# Polarization independent lattice-coupled terahertztoroidal excitations

ANGANA BHATTACHARYA[1],[*], BHAGWAT S. CHOUHAN,[1] BHAIROV BHOWMIK,[1] RAJAN SINGH,[2] AND GAGAN KUMAR,[1]

[1]*Department of Physics, IIT Guwahati, Guwahati, Assam, India-781039*
[2]*Centre for Nanotechnology, IIT Guwahati, Guwahati, Assam, India-781039*
*\*Corresponding author: angana18@iitg.ac.in*

**Abstract:** The toroidal dipole excitation is an important field for metamaterial research because of their low-loss attribute. In this study, we demonstrate numerically and experimentally, a unique polarization independent terahertz metamaterial that modulates a broad resonance into a sharp mode by coupling the inherent toroidal dipole excitation to the lattice mode of the metasurface. The advantage of polarization independence in the metasurface enables the excitation of "lattice-coupled toroidal mode" for both the transverse electric and transverse magnetic modes of the incident terahertz radiation. The interaction of the two dark low-loss modes results in the significant enhancement of the quality factor of the metasurface at the point of resonance matching. Such a polarization independent lattice-matched toroidal excitation- based device has the potential to impact the development of low loss terahertz component for ultrasensitive sensors, low loss equipment, and slow light devices for enhanced light matter interaction.

The terahertz (THz) range of frequencies have been widely explored in recent times because of their enormous potential in the development of high speed devices. The curiosity on THz technologies and its wide advantages led to increasing interest in its utilization [1] . Latest studies employing topological photonics to enable on-chip terahertz communication has received significant attention [2]. The advent of metamaterial (MM) research has boosted the growth of THz research. MMs are artificially designed materials constituting of periodic arrays of subwavelength resonators as its unit cells known as meta-atoms. THz metamaterials have been utilised to demonstrate properties that are not usually seen in natural materials, including electromagnetically induced transparency (EIT), slow light effect, cloaking, superlenses etc [3-5]. MMs have enabled the study of electromagnetic excitations that would otherwise have been difficult through natural materials. Toroidal excitation is one such type of electromagnetic excitation. The toroidal dipole moment is formed when magnetic dipoles in a material are aligned in a head-to-tail fashion. The alignment of magnetic moments leads to the excitation of the toroidal dipole resonance [6]. Toroidal resonances were studied to demonstrate narrow linewidths with high quality factors leading to reduction in radiative losses were significantly reduced in the dark toroidal modes. Several research groups have reported the analysis of toroidal dipole mode for use in sensors, switches, broadband modulators, biological sensors, multiband transparency devices, active devices [7-10]. Scientists have long sought to reduce radiative losses in the terahertz range. Besides the exotic toroidal resonance, another avenue for reduced radiation losses is the lattice mode of metasurfaces which depend on the periodicity of the MM. The lattice mode is a dark mode that arises due to diffraction along the interface of the periodically arranged meta-atoms of a metasurface. The energy transmitted is strongly confined to the surface, acting similar to a trap, hence reducing the far-field radiation [11]. The periodicity of the MM controls the resonant frequency of the lattice mode. Hence, the MM resonance can be easily coupled to the lattice mode by change in the periodicity without significant modification to the MM geometry. Coupling the MM resonance to the lattice mode has led to extremely high narrowing of resonance widths leading to very high-quality factor (Q) resonances and lowered radiative losses [12, 13] [14] [15]. Though several studies have reported the tuning of electric dipole and Fano resonance via lattice mode coupling, and the excitation of lattice induced transparency in metasurfaces, there has been a gap in the study of the coupling between toroidal resonance to the lattice mode [16]. Further, polarization dependence of the incident radiation to the lattice mode coupling becomes a setback in the device functionality. The coupling of the unique toroidal resonance to the lattice mode provides an interesting aspect in the dark-dark nature of their interaction.

In this letter we report, numerically and experimentally, the polarization independent coupling between a toroidal excitation to the first-order lattice mode (FOLM) in a planar terahertz metasurface leading to a sharp, high quality factor resonance. The unique MM geometry results in the reduction of resonance line-width, and an increase in quality factor of the resonance for both the TE and TM modes of the incident terahertz radiation. Such MM based devices that couple the dark lattice mode to the low-loss toroidal mode, independent of the periodicity, could be significant in the design of terahertz sensors and modulators. The polarization independence provides an extra degree of flexibility in device applications.

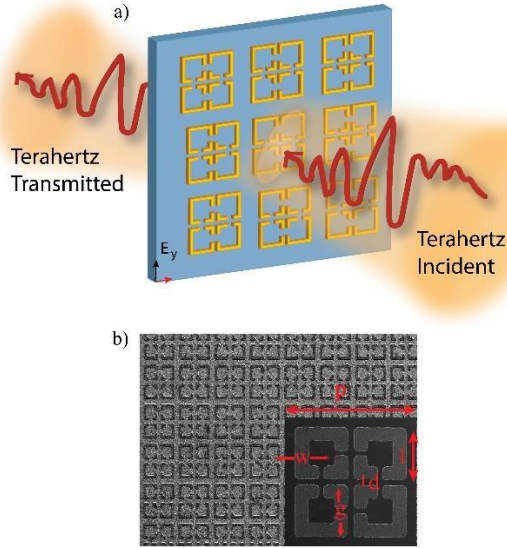

Figure 1: a) Schematic of the proposed metamaterial array having y-polarized terahertz radiation incident normally. b) SEM image of the fabricated metasurface array for P = 97 μm. Inset shows magnified view of the meta-molecule with periodicity 'p', length of each split ring resonator 'l', width 'w', capacitive gap 'g', and the distance between adjacent resonators 'd'.

Figure 1 (a) shows a schematic of the proposed MM array. Our proposed metamolecule has a simple symmetric geometry consisting of four Aluminum split-ring resonators (SRR) of width 'w' = 6 μm, length 'l' = 30 μm, capacitive gap 'g'= 6 μm and the distance between adjacent resonators, 'd' = 6 μm as shown in magnified image in figure 1(b). The metamolecules are periodically arranged in a silicon substrate of permittivity ε = 11.9 with initial periodicity 'p' = 97 μm. Further simulations were performed for a wide range of 'p'. The simulations were performed using CST microwave studio simulation software using tetrahedral meshing in the frequency domain. The incident terahertz radiation was simulated such that the electric field was polarized along the y direction initially. The cross polarized component is the one where the electric field is polarized along the x direction. The fabrication of the metamaterial samples was done using photolithography. Silicon substrate of 500 μm thickness initially deposited with an aluminium layer of thickness 200 nm. It was then spin-coated with S1813 positive photoresist followed by UV exposure using a hard mask. After development, it was etched using an aluminium etchant to obtain the desired MM sample. Each MM sample consists of a 1-centimetre by 1-centimetre array of the designed metamolecule arranged periodically for varying 'p'. Samples were fabricated for 'p' = 120 μm, 97 μm, and 75 μm and the THz transmittance through the samples was measured experimentally. An SEM image of the fabricated MM sample corresponding to P = 97 μm is shown in figure 1(b). The terahertz transmittance measurements were done using a switch based THz Time-domain spectroscopy setup. The red line in figure 2 (a) shows the transmissions spectrum for P = 97 μm of the MM. A sharp resonance is observed at 0.89 THz. The surface current profile at 0.89 THz is evaluated and clockwise current flow was observed in the left resonators while anti-clockwise current flow was observed in the right hand resonators of the meta-molecule. This indicates a head-to-tail formation of magnetic moments leading to a toroidal dipole excitation along the negative Y axis. This is shown in figure 3. Further, we study the lattice-coupled behaviour of the toroidal excitation at 0.89 THz. The lattice mode $F_L$, at normal incidence, is given by the formula $f_L = \frac{c\sqrt{i^2+j^2}}{nP}$, where c is the speed of light, n is the refractive index of the substrate or the medium of propagation of the lattice mode, and P is the periodicity of the MM. (i, j) indicate non-negative integers that define the order of the lattice mode [13]. The FOLM has an (i, j) value of (0,1). Hence, the FOLM frequency is given by, $f_L = \frac{c}{nP}$. Thus, the coupling to the FOLM can be achieved by tuning the periodicity of the metasurface. It has been reported that the coupling of a MM resonance to the FOLM leads to sharp resonant features with high quality factor. To verify this, we matched the $f_{FOLM}$ to the toroidal resonant frequency of the MM, i.e., 0.89 THz and evaluated the corresponding periodicity, P, which could excite the lattice mode at the selected frequency. Setting the values of $f_{FOLM}$=0.89 THz, c = 3 x $10^8$ m/s, and n = $\sqrt{\varepsilon}$ = $\sqrt{11.9}$, we obtained, P = 97 μm. Thus, the sharp toroidal mode was indeed verified to be the FOLM. To verify that the coupling between the toroidal mode and the FOLM leads to line-width reduction in the MM, we measured the THz transmittance for varying periodicities P = 120 μm, and P = 75 μm of the metasurface. The results obtained from the experimental measurements are shown in figure 2(a) while the numerically simulated results are shown in figure 2 (b).

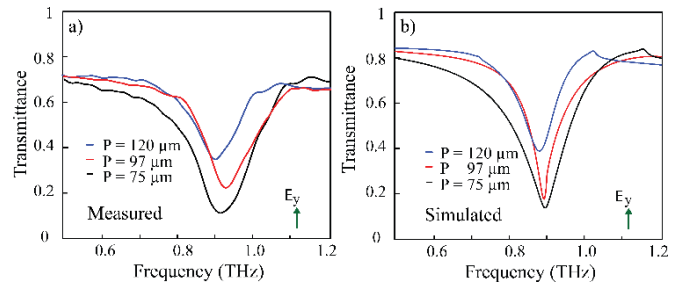

Figure 2: a) Measured transmittance spectra for changing periodicities 'P' for y-polarized THz light. b) Simulated transmittance spectra for varying periodicities of the MM for y-polarized incident light.

The sharp red curves in figures 2 (a,b) depicts the transmittance for P = 97 μm. The black lines in figures 2 (a,b) indicates the transmittance for P = 75 μm and it was observed that the linewidth at 0.89 THz increased and a resonance broader than the one for P = 97 μm was excited. Further, the transmittance for P = 120 μm was evaluated as shown by the blue lines in figures 2(a,b). A broad resonance was observed at 0.89 THz for P = 120 μm with around 50% transmittance amplitude. We observed that the measured transmittance spectrum for P = 120 μm, 97 μm, and 75 μm matched closely with the simulation results. A slight shift in the resonance frequency was observed which may be attributed to human errors in the fabrication process. Further, the measured transmission spectrum does not reflect the sharpness of the lattice coupled mode at P = 97 µm, which was believed to be a result of the limitations in the resolution of the experimental THZ-TDS setup.

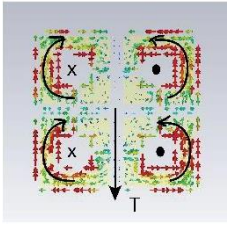

*Figure 3: Surface current profile for P = 97 µm in the MM. The arrows indicate the direction of flow of the surface current.*

From these observations, it was verified that modulation of the transmittance amplitude and resonance sharpness by tuning of periodicity, and the reduced line-width of the resonance at P= 97 µm, was a result of the coupling of the toroidal mode with the FOLM. Hence, coupling to of the FOLM to the toroidal mode resulted in a sharp "toroidal-coupled lattice mode" with increased transmission depth. The decrease in linewidth of the toroidal excitation is further quantified by evaluating the quality factor (Q) of the resonance. The quality factor (Q) of a resonant circuit is a dimensionless parameter defined as ratio of the energy stored in the mode to the energy dissipated per cycle of oscillation. A high Q indicates low loss. We used the Fano line shape fitting formula to evaluate the Q. The formula is given by, $T = (a + ib + \frac{c}{(\omega-\omega_0+i\gamma)})^2$, where T is the transmittance and a, b, c are constants. '$\omega$' is the frequency range while $\omega_0$ is the resonance frequency [17]. '$\gamma$' indicates the damping rate in the circuit. The Q value is calculated by, $Q = \omega_0/2\gamma$. In this study, we Fano-fitted the transmittance for different values of periodicity of the metasurface and reported the resultant Q of the toroidal mode. Figure 3(a) depicts the variation of Q with changing periodicities.

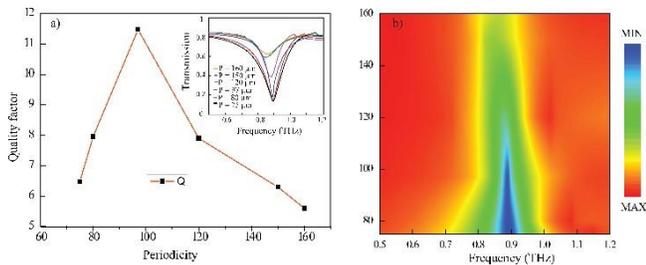

*Figure 4: a) Quality factor variation of the toroidal resonance at 0.89 THz for different values of periodicity 'P' of the MM. b) Modulation of the transmittance line-width for varying P of the MM.*

The Q for P = 97 µm, i.e, the lattice coupled toroidal was evaluated to be equal to 11. 47. Further, the Q for P = 75 µm was evaluated to 6.48, while that for P = 120 µm evaluated to 7.9. The inset of figure 3 (a) depicts the tuning of resonance line-width on variation of periodicity. It may be observed that there is a sharp rise in the Q for the lattice-coupled toroidal mode at P = 97 µm. As the periodicity is increased to beyond 120 µm the Q values falls offs and nears about 5.8 for P = 160 µm and higher. This matches the transmittance shown in the inset of figure 3(a) where it is evident that increasing the periodicity leads to a broad resonance whose line-width becomes almost constant for periodicities far from the FOLM. Hence, as predicted, the coupling of toroidal mode to the FOLM resulted in a high Q, sharp resonance mode. A 51% increase in Q is observed for the lattice-coupled toroidal mode for P = 97 µm as compared to the Q at higher periodicity of 160 µm. Moreover, 44% increase in Q is reported as compared to the lower periodicity, P = 75 µm. The contour plot in figure 3 (b) depicts line-width of the resonance for varying P in the frequency range of 0.5 THz to 1.2 THz.

In order to provide an extra degree of freedom in applications of the toroidal-coupled lattice mode for terahertz meta-devices, we explored the polarization independent characteristics of the proposed MM. Initially the transmittance was reported for y polarized incident THz, radiation falling normally on the MM sample. Next, measurements are performed to study the toroidal mode, such that the electric field is polarized along the x direction. Simulations are performed to verify the experimental results. The experimental measurements are shown in figure 5 (a) while figure 5 (b) shows the simulated results.

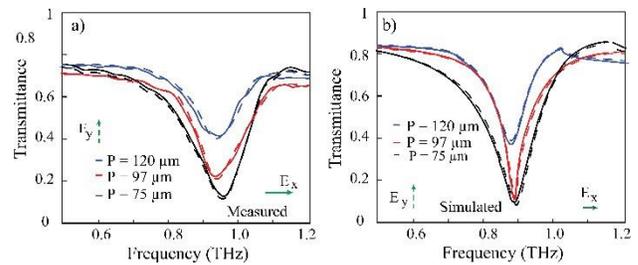

*Figure 5: a) Measured transmittance for varying periodicity when electric field of incident THz radiation is polarized along the x-axis. The dotted lines indicate the simulated transmittance for the for y-polarized THz radiation. b) Simulated transmittance for varying periodicity of MM for x-polarization of incident THz radiation. The dotted lines show measured transmittance for y-polarized THz beam.*

The blue line in figure 5 (a) depicts the measured x-polarized transmittance for P = 120 µm, the red line depicts that for P = 75 µm, while the black curve depicts the same for P = 97 µm. The dotted lines indicate the corresponding transmittance for the y-polarized THz radiation that was measured and simulated previously. It may be observed from figure 5 (a) that the x-polarized transmittance matches exactly with the simulated results for the y-polarized transmittance. The lattice coupled mode is observed at 0.89 THz for P = 97 µm. It can be observed from figure 5 (b) that the simulated results closely match with the measured results for the x-polarized mode. The blue curve, red curve, and black curve in figure 5(b) depict the simulated transmittance for P = 120 µm, P = 97 µm, and P = 75 µm respectively. Thus, it is experimentally and numerically verified that the toroidal-coupled lattice mode is invariant of the polarization of the incident THz radiation.

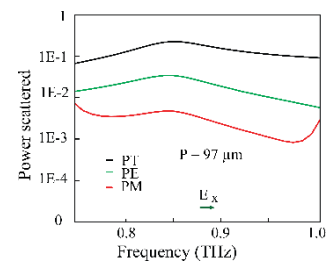

*Figure 6: Multipole analysis of the power scattered by electric, magnetic, and toroidal dipole moments.*

Further, the nature of the resonance at 0.89 THz for the x-polarized incident THz radiation is analyzed by performing a multipole anlaysis of the scattered powers of radiation by the electric, magnetic, and toroidal dipole moment [18]. The multipole analysis is shown in figure 6. The plan line depicts the power scattered by toroidal moment, the green line shows the electric scattered power while the red line shows the scattered power by magnetic moment. The analysis is done for the lattice coupled mode at P = 97 μm. It is observed that the toroidal mode has highest scattered power indicating that the resonance at 0.89 THz for the x-polarized beam is a lattice-coupled toroidal mode.

In conclusion, we have successfully demonstrated the polarization independent coupling of the exotic toroidal resonance to a first-order lattice mode. Measurements for varying periodicities, for y-polarized THz radiation, demonstrated that the lattice coupled mode for periodicity P = 97 μm at 0.89 THz shows decreased line-width of the resonance. Surface current profile indicated that the MM geometry has an inherent toroidal dipole excitation at 0.89 THz. The quality factor (Q) calculation of the lattice-coupled toroidal mode shows a 51% increase in Q for the metasurface configuration corresponding to P = 97 μm as compared to higher periodicities. Further, the transmittance for the x-polarized THz radiation is measured, and it is reported that the metasurface demonstrates polarization independent lattice coupled toroidal mode. Multipole analysis of the scattered power of electromagnetic radiation confirmed toroidal domination for the lattice-coupled mode for x-polarized light. Thus, we have demonstrated a terahertz meta-device that excites a polarization independent sharp toroidal-coupled lattice mode. To our knowledge, this is the first demonstration of a polarization independent lattice coupled toroidal excitation in metasurfaces. Such an increase in Q of a dark toroidal mode via coupling to the lattice mode can find wide applications in polarization independent highly sensitive sensors, modulators, and photonic devices for terahertz applications.

The authors would like to acknowledge the financial support from Science and Engineering Research Board (CRG/2019/002807).